\documentclass[copyright]{eptcs}
\providecommand{\event}{UNIF 2010} 
\usepackage{breakurl}             

\usepackage[latin1]{inputenc}
\usepackage{epsfig, amssymb, amsmath}
\usepackage{latexsym, graphics, graphicx}
\usepackage{yfonts}
\usepackage{proof}
\usepackage{algorithmic}
\usepackage{algorithm}
\usepackage{amsthm} 
\usepackage{amsxtra}
\usepackage{epsfig}
\usepackage{color}
\usepackage{float}
\newcommand{\ignore}[1]{}

\newcommand{\eq}{\mathcal{EQ}}

\newcommand{\ueq}{=_{}^?}

\theoremstyle{plain}
\newtheorem{theorem}{Theorem}[section]
\newtheorem{Lemma}[theorem]{Lemma}

\theoremstyle{definition}
\newtheorem{definition}[theorem]{Definition}

\title{On the Complexity of the Tiden-Arnborg Algorithm for Unification
modulo One-Sided Distributivity}
\author{Paliath Narendran\footnote{Partially supported by the NSF grants
CNS-0831209 and CNS-0905286}
\institute{University at Albany--SUNY\\
College of Computing and Information\\
Computer Science Department}
\email{dran@cs.albany.edu }
\and Andrew Marshall\footnote{Partially supported by the NSF grants
CNS-0831209 and CNS-0905286}
\institute{University at Albany--SUNY\\
College of Computing and Information\\
Computer Science Department}
\email{marshall@cs.albany.edu  }
\and Bibhu Mahapatra
\institute{New York State\\
 Education Department}
\email{bmahapat@mail.nysed.gov}
}
\def\titlerunning{The Complexity of the Tiden-Arnborg Algorithm for Unification
modulo One-Sided Distributivity}
\def\authorrunning{P. Narendran, A. Marshall \& B. Mahapatra}
\begin{document}
\maketitle

\begin{abstract}
We prove that the Tiden and Arnborg algorithm for equational unification modulo {\em one-sided\/} distributivity
is not polynomial time bounded as previously thought. A set of counterexamples is developed that demonstrates 
that the algorithm goes through exponentially many steps.
\end{abstract}

\section{Introduction}
Equational unification is central to automated deduction and its applications 
in areas such as symbolic protocol analysis. In particular,
the unification problem for the theory $AC$ 
(``Asso\-ciativity-Commu\-tativity'') and its extensions $ACI$ (``AC plus Idempotence'')
and $ACUI$ (``ACI with Unit element'') have been studied in great
detail in the past. Distributivity (of one binary operator over another) has received
less attention comparatively. Some significant results have been
obtained such as Schmidt-Schauss' breakthrough decidability 
result~\cite{Schmidt-Schauss98} for unification modulo
the theory of two-sided distributivity \begin{eqnarray*}
x \times (y + z) & = & (x \times y) + (x \times z) \\
(y + z) \times x & = & (y \times x) + (z \times x) \end{eqnarray*}
Other works include~\cite{DBLP:journals/jsc/Contejean93,DBLP:conf/icalp/Contejean93}.

One of the earliest papers that considered a subproblem of this
is by 
Tiden and Arnborg~\cite{TidenArnborg87}. They present an algorithm for equational 
unification
modulo a {\em one-sided\/} distributivity axiom: \[ x \times (y + z) = x
\times y + x \times z \] This unification problem has recently been of interest
in cryptographic protocol analysis since many cryptographic
operators satisfy this property: for instance, modular exponentiation
(used in the RSA and El~Gamal public key algorithms) distributes over
modular multiplication. Indeed, many electronic election protocols rely on the
property of ``homomorphic encryption'' where encryption distributes over
some other operator. (A new algorithm for this unification
problem, using a novel approach, is given
in~\cite{HaiLinthesis,ALLNR-capunif}.)

Our goal in this paper is to analyze the Tiden-Arnborg algorithm.
We prove that the algorithm 
is not polynomial time bounded as claimed in the Tiden-Arnborg paper. 
A set of counter examples is outlined that demonstrates that
the present algorithm goes through exponentially
many steps.


\subsection{The Tiden-Arnborg Algorithm}

We present a very brief description of the algorithm of Tiden and
Arnborg using deduction (inference) rules. 
First of all, it should be pointed out that what they consider
is the {\em elementary\/} unification 
problem~\cite{BaaderSnyder}, where the terms can only
contain symbols in the signature of the theory and variables. (Thus free
constants and free function symbols are not allowed.) Hence
we can assume without loss of generality that the input
is given as a set of equations where each equation is in one of
the following forms: \[ X =_{}^? Y, ~ X =_{}^? Y + Z, ~
\mathrm{and} ~ X =_{}^? Y \times Z \] The key steps in the algorithm
can be described by the following deduction rules:\\[-5pt]

\begin{tabular}{lcc}
(a) & $\qquad$ & $\vcenter{
\infer[\qquad \mathrm{if} ~ U ~ \mathrm{occurs ~ in} ~ \eq ]{\{U =_{}^? V\} \cup \, [V/U](\eq) }
      { \{U =_{}^? V\} ~ \uplus ~ \eq }
}
$\\[+30pt]
(b) & & $\vcenter{
\infer{\eq ~ \cup ~ \{ U =_{}^? V \times W, \; V =_{}^? X, \; W =_{}^? Y \}}
{\eq ~ \uplus ~ \{ U =_{}^? V \times W, \; U =_{}^? X \times Y \}}
}
$\\[+30pt]
(c) & & $\vcenter{
\infer{\eq ~ \cup ~ \{ U =_{}^? V + W, \; V =_{}^? X, \; W =_{}^? Y \}}
{\eq ~ \uplus ~ \{ U =_{}^? V + W, \; U =_{}^? X + Y \}}
}
$\\[+30pt]
(d) & & $\vcenter{
\infer{\eq ~ \cup ~ \{ U =_{}^? V \times W, \; W =_{}^? W_1^{} + W_2^{}, \; X =_{}^? V \times W_1^{}, \; Y =_{}^? V \times W_2^{} \}}
{\eq ~ \uplus ~ \{ U =_{}^? V \times W, \; U =_{}^? X + Y \}}
}
$
\end{tabular}

\vspace{0.2in}

\vspace*{0.8em}
The $W_1, W_2$ in rule~(d) are fresh
variables and $\uplus$ is
{\em disjoint union}. Furthermore, rule~(d) 
(the ``splitting rule'') is applied only when the
other rules cannot be applied. A set of equations is said to be {\em simple\/}
if and only if none of the rules (a), (b) and (c) can be applied to it.
In other words, in a simple system, no variable can occur as the 
left-hand side in more than two
equations.
A {\em sum transformation\/} is defined as a binary relation between two simple
systems $S_1^{}$ and $S_2^{}$, where $S_2^{}$ is obtained 
from $S_1^{}$
by applying rule~(d),
followed by repeated exhaustive applications of rules (a), (b) and (c). Clearly,
a sum transformation is applicable if and only if some variable occurs 
as the left-hand side in more than one
equation.

Detection of failure is done using a kind of ``extended
occur-check'' using two graph based data structures. 
We repeat the definitions of the graph structures and give a 
sketch of the algorithm presented in Tiden and Arnborg~\cite{TidenArnborg87} 
for the convenience of the reader.   
\vspace{0.1in}
\begin{definition}
 The {\em dependency graph\/} of a simple system, $\varSigma$, is an 
edge colored, directed multi-graph.
It has as vertices the variables of $\varSigma$. For an equation $x= y + z$ in $\varSigma$
it has an $l_{+}$-colored edge $(x,y)$ and an $r_{+}$-colored edge $(x,z)$. An equation
$x = y \times z$ similarly generates two edges with colors $l_{\times}$ and $r_{\times}$.\\
\end{definition}

\begin{definition}
The {\em sum propagation graph\/} of a simple system $\varSigma$ is a
directed simple graph.  It has as vertices the equivalence classes of
the symmetric, reflexive, and transitive closure of the relation
defined by the $r_{\times}$-edges in the dependency graph of
$\varSigma$.  It has an edge $(V,W)$ iff there is an edge in the
dependency graph from a vertex in $V$, to a vertex in $W$ with color
$l_{+}$ or $r_{+}$.
\end{definition}

The dependency graph structure is sufficient for finding all the
occur-check like errors that may develop as the algorithm works
with the system of equations. 
The propagation graph is needed to detect non-unifiable systems that
cause infinitely many applications of the splitting rule~(d).
An
example of this type of system is the following two
equations:
\begin{center}
 $Z =^? V_2 + V_3$,  $Z =^? V_1 \times V_3$.
\end{center}
These types of systems are shown not to have a unifier and as
they will never produce a cycle in the dependency graph, the
propagation graph is needed.

Tiden and Arnborg give a polynomial time procedure for producing a
simple system of equations form an initial set of equations. We sketch
their unification algorithm from the starting point of an initial
simple system.

\begin{algorithm}[H]
\caption{UNIFY~\cite{TidenArnborg87}}
\label{alg1}
\begin{algorithmic}
\REQUIRE Simple system $\varSigma_1$.\\
 $k:=1$
\WHILE{ The sum transformation can be applied}
\item If either the dependency or propagation graph contains a cycle, then stop with failure.
\item Using the sum transformation compute $\varSigma_{k+1}$
\item $k := k+1$
\ENDWHILE
\item Compute the most general unifier ($mgu$) by back substitution.
\end{algorithmic}
\end{algorithm}

It is shown that if a system
is not unifiable it
will, after finitely many applications of the sum transformation, produce a
cycle in one of the graphs. It is also shown that if a system
is unifiable then the algorithm will produce the
$mgu$.

In the next section we present a family of unifiable systems that
produce no cycles in either graph, but require exponentially many
applications of the sum transformation.


\section{Counterexamples}

We present a family of \textit{unifiable} simple systems on which 
the Tiden-Arnborg
algorithm runs in exponential time. For ease of exposition, we
only use the letters $T$,
$x$ and $y$ for variables, along with subscripts for $x$ and $y$
which are strings over the alphabet $\{ 1, 2 \}$.

\begin{definition} \label{EQ}
 Let EQ be a subset of the simple system defined as follows: all
 multiplications are of the form $x_{i} \ueq T \times y_{j}$ (or
 $y_{j}  \ueq  T \times x_{i}$) where $T$ is a unique variable and all
 additions are of the form $x_{i} \ueq x_{i1}+x_{i2}$ or
 $y_{i} \ueq y_{i1}+y_{i2}$.
\end{definition}
As the left variable of the multiplication operation will not effect the complexity result 
we use the unique variable $T$ in this position. This makes the proof simpler. 
Thus the splitting rule~(d) above can be viewed as\\[-10pt]

\begin{tabular}{lcc}
 & & $\vcenter{
\infer{\eq ~ \cup ~ \{ U_i =_{}^? T \times W_j, \; W_j =_{}^? W_{j1}^{} + W_{j2}^{}, \; U_{i1}^{} =_{}^? T \times W_{j1}^{}, \; U_{i2}^{} =_{}^? T \times W_{j2}^{} \}}
{\eq ~ \uplus ~ \{ U_i =_{}^? T \times W_j, \; U_i =_{}^? U_{i1}^{} + U_{i2}^{} \}}
}
$
\end{tabular}

\noindent
where $U, W \, \in \, \{ x, y \}$.\\
Specifically, we examine the complexity of unifying a set of
equations from EQ. It will be shown that to achieve a unifier, 
the Tiden-Arnborg algorithm requires exponentially many steps.

\begin{definition} \label{sigman}
For $n \ge 0$, let $\sigma(n)$ be the set of equations
\begin{eqnarray*}
x_{1^{i}} &  \ueq  & x_{1^{i+1}}+x_{1^{i}2}, \\
y_{2^{i}} &  \ueq  & y_{2^{i}1}+y_{2^{i+1}}, \\
y_{2^{i}1} &  \ueq  & T \times x_{1^{i}2},\\
x &  \ueq  & T \times y, \\
x_{1^{i+1}} &  \ueq  & x_{1^{i+2}}+x_{1^{i+1}2}
\end{eqnarray*}
for all $0 \le i \le n$.
\end{definition}

Thus $\sigma(0)$ is $\{ x  \ueq  x_1 + x_2, \, y  \ueq  y_1 + y_2, \,
x_{1}  \ueq  x_{11} + x_{12}, \,    
x   \ueq   T \times y, \,
y_{1}  \ueq  T \times x_{2}
\}$.\\

Similarly $\sigma(2)$ is $\{ x  \ueq  x_1 + x_2, \, y  \ueq  y_1 + y_2, \,
x_{1}  \ueq  x_{11} + x_{12}, \,    y_{2}  \ueq  y_{21} + y_{22}, \,
x_{11}  \ueq  x_{111} + x_{112}, \, y_{22}  \ueq  y_{221} + y_{222}, \, 
x_{111}  \ueq  x_{1111} + x_{1112}, \, 
x   \ueq   T \times y, \,
y_{1}  \ueq  T \times x_{2}, \,
y_{21}  \ueq  T \times x_{12}, \,
y_{221}  \ueq   T \times x_{112} \}$.\\

Note that $\sigma(k + 1)$ =  $\sigma(k)$ $\cup$
$\{ y_{2^{k+1}}   \ueq   y_{2^{k+1}1}+y_{2^{k+2}}, \,
    y_{2^{k+1}1}  \ueq   T \times x_{1^{k+1}2}, \,
    x_{1^{k+2}}   \ueq   x_{1^{k+3}}+x_{1^{k+2}2} \}$
for all $k \ge 0$. \\

\begin{definition} \label{peak}
 We denote a variable $x_{i}$ (or $y_{i}$) as a
 \textit{peak} iff there are equations 
 $x_{i} \ueq x_{i1}+x_{i2} \text{ and } x_{i} \ueq  T \times y_{j}$ 
 (or $y_{i} \ueq y_{i1}+y_{i2} \text{ and } y_{i} \ueq 
 T \times x_{j} $)
\end{definition}

We claim that a system of equations, as defined in
Definition~\ref{sigman}, will result in exponentially many applications of
the sum transformation rule.

\section{Proof}

For a set of equations $S$, 
let $m(S)$ denote the number of $\times$ symbols in it and
$p(S)$ denote the number of + symbols in it.
Consider the sets of equations defined in 
Definition~\ref{sigman}. 
By the analysis in~\cite{TidenArnborg87} the number of sum transformations
should be bounded by $ m(S) * p(S) $. 
We can see that according to Definition~\ref{sigman}
$m (\sigma(n)) = n + 2$ and $p (\sigma(n)) = 2n + 3$.
Thus the upper bound should be $2n_{}^2 + 7n + 6$. However,
the actual bound for systems of equations $\sigma (n)$
will be shown to be $2^{n+3}-(n+4)$. 

\begin{figure}[h] 
  \centering
   \input{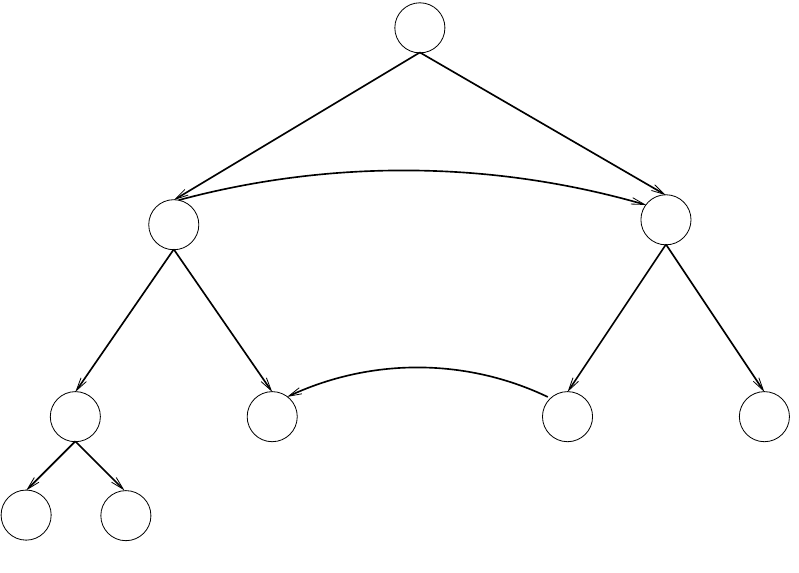}
  \caption{Example graph with a peak at node $x$} \label{fig1}
\end{figure}

We can view the sets of equations defined in Definition~\ref{sigman} 
as tree-like
graphs. Nodes correspond to variables.
We first add a dummy root node with outdegree~2 whose children are
the 
initial nodes $x$ and $y$. 
The summation equations are represented by downward
edges, from every parent node to its two children.
We represent the multiplication equations as lateral edges,
i.e., edges between
nodes at the same {\em level,\/} i.e., distance
from the root node. (Thus the graph is not really a tree
if lateral edges are considered.). 
Because all left multiplication edges goto $T$ and have no effect
on the complexity of the algorithm in these systems
of equations, we leave these edges out of the diagrams for
clarity.  Let $G (n)$ be the graph of $\sigma (n)$
See Figure~\ref{fig1} for $G (0)$.
Note that the height of the tree is 3, i.e, there are 3
levels. In general, the graph of $\sigma (n)$ has $n+3$ levels.
We view the algorithm as proceeding down the
tree, with sum transformations at a level completed before starting at the 
next level. We analyze the complexity of the Tiden-Arnborg algorithm
in terms of transformations done on the graph as the algorithm proceeds.
We show that if $l$ is the height of the tree, then the number of sum transformations
applied is $2^l - (l + 1)$.
 
\begin{figure}[h]
\begin{center}
\input{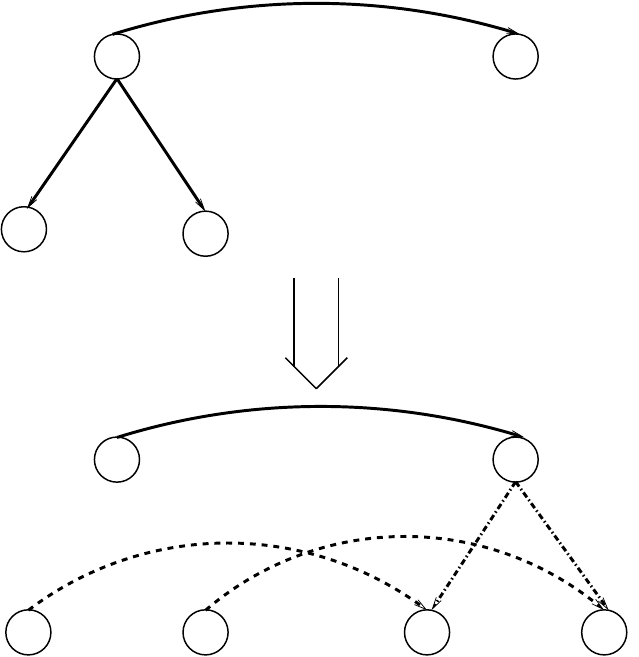}
\end{center}
\caption{Sum transformation} \label{sumtrans}
\end{figure}

Observe that a variable is a peak if and only if its node has both 
downward and lateral edges. Figure~\ref{sumtrans} shows the effect of
a sum transformation at a peak on the graph. Note that lateral edges 
are never deleted. Each application of the sum transformation 
increases the number of lateral edges by at most~2.

Note also that other than at the lowest
level (depth $n+3$) the graph will have initially one multiplication or
exactly one edge between nodes at the same height in the graph. We
can also see that the graph is partitioned between the left and
right side or $x$ and $y$ side and that at level one, there is an
edge from $x$ to $y$. However, at all other lower levels, the
initial edge between nodes of the same level goes from $y$ to $x$.

\begin{figure}[h]
  \centering
   \input{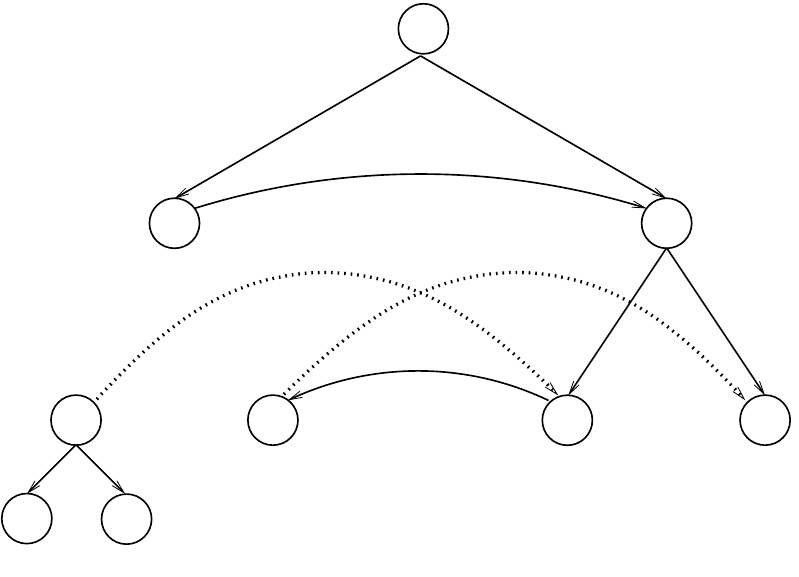}
  \caption{Graph after one application of the sum transformation at node $x$, creating a new peak at node $x_{1}$ and new edges from $x_{1}$ to $y_{1}$ and $x_{2}$ to $y_{2}$} \label{fig2}
\end{figure}

We can also see that given the graph as described above, each
time the sum transformation is applied, the peak moves from
either the $x$ side of the graph to the $y$ side or from the $y$
side to the $x$ side, and the new peak was not previously a
peak. To see this, take any system as defined by
Definition~\ref{sigman} and examine the graph of that
system. Initially all edges from nodes at the same height only go
from one side to the other. In this limited formulation of
Definition~\ref{sigman} these same level edges are the only
multiplication functions. This ensures that any time a sum
transformation is performed on some equation, $x_{i} \ueq T \times
y_{j}$ or $y_{i} \ueq T \times x_{j}$, by definition the new edges
created by the sum transformation must go from either $x$ to $y$
or $y$ to $x$ because there are no multiplication equations of
the form $x_{i} \ueq  T \times x_{j}$. The fact that a new peak was
not previosly a peak follows from Definition~\ref{sigman} and the
definition of the sum transformation. Since we assume a simple
system of equations, there are no two distinct equations of the
form $x \ueq x_{i} + x_{j}, ~ x \ueq x_{k} + x_{l}$ (i.e., with the same
variable on the left-hand side): likewise for $y$. Once when the
sum transformation is applied to equations $x_{i} \ueq T \times y_{j}$
and $x_{i} \ueq x_{i1} + x_{i2}$, the downward edges from $x_{i}$ to
both $x_{i1}$ and $x_{i2}$ are removed (see Figure~\ref{fig2}). 
Thus $x_{i}$ can never become a peak again.
\begin{Lemma} \label{Lemma1}
Sum transformations at level $k$ create a peak at $x_{1^k}^{}$ at level
$k+1$ provided $k+1$ is not the lowest level.
\end{Lemma}
\begin{proof}
 This follows inductively from the form of the graph in
 Definition~\ref{sigman}. First from the definition of the set of
 equations, the first peak is located at $x$. After the first sum
 transformation a peak is created at $x_{1}$. Assume this
 propagates to level $n$. Then there is a peak at $x_{1^{n}}^{}$ to
 which the sum transformation is applied, adding an equation of
 the form $x_{1^{n+1}} \ueq T \times y_{1^{n+1}}$, which creates a peak
 at the next level, provided there is already an equation
$x_{1^{i+1}}  \ueq   x_{1^{i+2}}+x_{1^{i+1}2}$.
\end{proof}

\begin{Lemma} \label{sinknode}
 There are no lateral edges from nodes corresponding to variables of the form
$y_{2^j}^{}$ for $j \ge 0$. In other words, no equations of the
form $y_{2^j}^{}  \ueq  T \times x_k^{}$ are generated.
\end{Lemma}
\begin{proof}
This follows inductively from Definition~\ref{sigman}. In any initial
 system there is no edge from any $y_{2^k}$ node at level $k$. By
 the definition of sum transformation an outgoing lateral edge from $y_{2^k}$
 can be created only if the parent node, $y_{2^{k-1}}$ has an outgoing lateral edge.
\end{proof}

We can also notice a fact about the order in which the nodes
become peaks via the sum transformation. The
order is a right-to-left lexicographical order of the digits of
the nodes' indices (i.e., subscripts). 
For example, for level 4 the sequence is
$x_{111}\rightarrow y_{111} \rightarrow x_{211} \rightarrow
y_{211} \rightarrow x_{121} \rightarrow y_{121} \rightarrow
x_{221} \rightarrow y_{221} \rightarrow x_{112} \rightarrow
y_{112} \rightarrow x_{212} \rightarrow y_{212} \rightarrow
x_{122} \rightarrow y_{122} \rightarrow x_{222} \rightarrow
y_{222}$. Note, $y_{222}$ is not necessarily a peak but is added to illustrate the path. Based on this observation we have the following lemma:
\begin{Lemma} \label{Lemma2}
At any level of the tree, if
$x_{i} \ueq T \times y_{j} $ is an equation (i.e., if there is a
lateral edge from $x_i$ to $y_j$) then $i = j$. Similarly,
if $y_{i} \ueq T \times x_{j}$ is an equation, then
$j=revlex(i)$ where {\em revlex\/} is the lexicographic
successor of the index of the node, $y_i$, but starting with the
$1^{st}$ bit (i.e., from right to left).
\end{Lemma}
\begin{proof}
 This follows inductively from Definition~\ref{sigman} and the sum
 transformation. The base cases are $x  \ueq  T \times y$ (level~1) and
 $y_{1} \ueq T \times x_{2}$ (level~2). Assume this property for level $k$. Now
we will show that all the equations introduced at level $k + 1$
by sum  transformation at level $k$ will satisfy the property.
If $y_{i} \ueq T \times x_{revlex(i)}$ is an equation at level $k$ and
$y_i  \ueq  y_{i1} + y_{i2} $ is an equation (i.e., $y_i$ is a peak at level $k$), 
then applying the sum
transformation results in 
$y_{i1} \ueq T \times x_{revlex(i)1} \text{ and } y_{i2} \ueq T \times
x_{revlex(i)2}$.
Now note that $revlex(i 1) =  {revlex(i)1}$
and $revlex(i 2) = {revlex(i) 2}$ since $i$ is not a string of 2's.
Note also that if $k$ is not the lowest level, then there will
already be an equation $y_{2^{k}1}   \ueq   T \times x_{1^{k}2}$ at level~$k+1$
but this does not violate the property in the lemma since
$revlex ( {2^{k}1} ) = {1^{k}2}$.

If $x_{i} \ueq T \times y_{i} \text{ and }
 x_{i} \ueq x_{i1} + x_{i2} $, then by application of the sum
 transformation $x_{i1} \ueq T \times y_{i1} \text{ and } x_{i2} \ueq T \times
 y_{i2}$ and the result follows.
\end{proof}

\begin{Lemma} \label{peaks-on-paths}
If there is a path of lateral edges from node $u_i$ to node $v_i$
in the graph at some point where node $u_i$ is a peak, then
every node on the path, except possibly $v_i$, will become a peak at some 
point.
\end{Lemma}
\begin{proof}
Straightforward, by induction on the length of the path.
\end{proof}

\begin{Lemma} \label{lemma6}
At every level $k < n + 3$ a path of lateral edges between
$x_{1^{k-1}}$ to $y_{2^{k-1}}$ is created.
\end{Lemma}
\begin{proof}
For brevity, we refer to such paths as RL paths. Clearly there (already) is
an RL path at level~1.
We show that if a 
RL path exists at level $k$ and $k+1 < n+3$,
then a RL path will be created at level 
$k+1$. By Lemma~\ref{Lemma1} there will be a peak at $x_{1^{k-1}}$ and by 
Lemma~\ref{peaks-on-paths} every node other than $y_{2^{k-1}}$ will become a peak.
This creates, at level $k+1$, edges of the form 
$x_{i1} \ueq T \times y_{i1}$, $x_{i2} \ueq T \times  y_{i2}$, 
$y_{i1} \ueq T \times x_{revlex(i)1}$ and 
$y_{i2} \ueq T \times x_{revlex(i)2}$ for every $i \neq 2^{k-1}$.
Since the edge $y_{2^{k-1}1}  \ueq   T \times x_{1^{k-1}2}$ is already
there to begin with, we get the RL path at level~$k+1$.
\end{proof}

\begin{Lemma} \label{Lemma3}
 At each level $k < n+3$ of the graph, the sum transformation can be
 applied $2^{k}-1$ times.
\end{Lemma}
\begin{proof}
 Follows from Lemma~\ref{lemma6}. At each level $k$, 
$2^{k}$ nodes will be created eventually. 
An RL path can be created and thus the 
sum transformation must be applied to all nodes except $y_{2^{k-1}}$ 
resulting in $2^{k}-1$ applications at each level $k$.
\end{proof}
\begin{theorem}\label{thm}
 For a graph of height $n$, $2^{n+1}-n-2$ sum transformations are used.
\end{theorem}
\begin{proof}
 This easily follows from Lemmas~\ref{Lemma1}--~\ref{Lemma3} and the fact that
 ${\sum}_{i=0}^{n}\left( 2^{i}-1\right) = 2^{n+1}-n-2$. 
\end{proof}
We see that the current algorithm fails to achive polynomial
complexity for at least a subset of possible unification
problems. Further counter examples may be found that also cause
this exponential growth with the sum transformation. This
naturally results in the question of whether a polynomial time
algorithm can be found, either by a modification of the current
algorithm or by a new approach.

\section{An Illustrated Example}
\begin{figure}[h]
  \centering
   \scalebox{1.5}{ \input{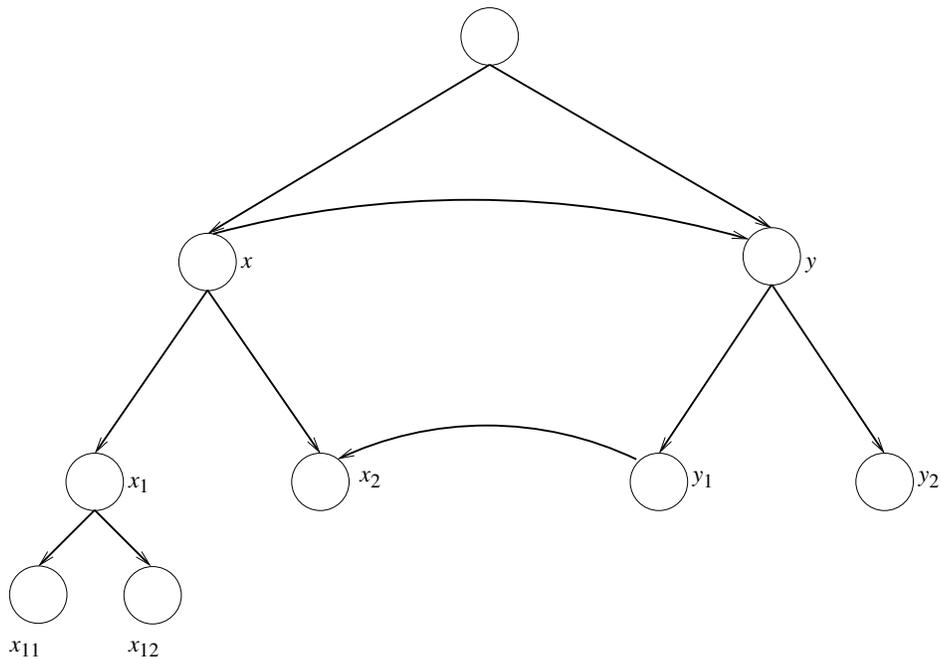} }
  \caption{Graph for $\sigma(0)$} \label{fig3}
\end{figure}

\begin{figure}[H]
  \centering
   \input{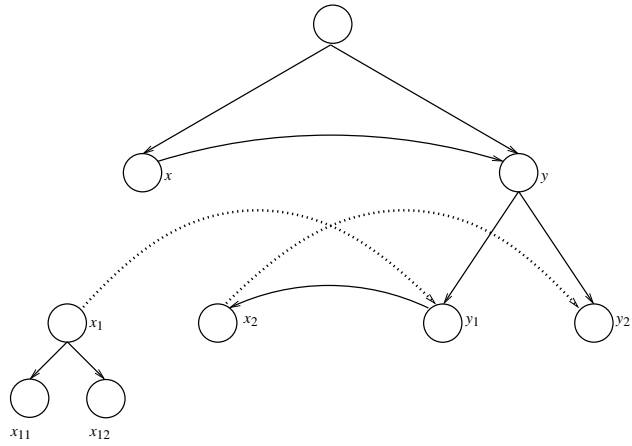}
  \caption{After one application of the sum transformation} \label{fig4}
\end{figure}

In this section we give an example of the process on a system of equations 
defined as in Definition~\ref{sigman}.
We begin with $\sigma(0)$, i.e., the following set of initial equations:
 \begin{eqnarray*}
 x & \ueq & T \times y,\\
 x & \ueq & x_{1}+x_{2},\\
 y & \ueq & y_{1}+y_{2},\\
 y_{1} & \ueq & T \times x_{2},\\
 x_{1} & \ueq & x_{11}+x_{12}\\
\end{eqnarray*}
This can be represented by a graph as shown in Figure~\ref{fig3}. 
Note that the first peak is located at node $x$.
The first peak, $x$, is selected and the sum transformation
can be applied, resulting in the removal of equation
$x \ueq x_{1}+x_{2}$ from the set of equations and the addition of the
two equations $x_{1} \ueq T \times y_{1}$ and $x_{2} \ueq T \times
y_{2}$. The direction of the new edges are from $x$ to $y$ due to
the fact that the multiplication equation from the peak, to which
the sum transformation was applied was also from $x$ to $y$. After the sum transformation is applied $x$
is no longer a peak because of the removal of $x \ueq x_{1}+x_{2}$,
but now the node $x_{1}$ is a peak due to the addition of
$x_{1} \ueq T \times y_{1}$ (see Lemma~\ref{Lemma3}). This new graph is shown
in Figure~\ref{fig4}.

\begin{figure}[H]
  \centering
   \input{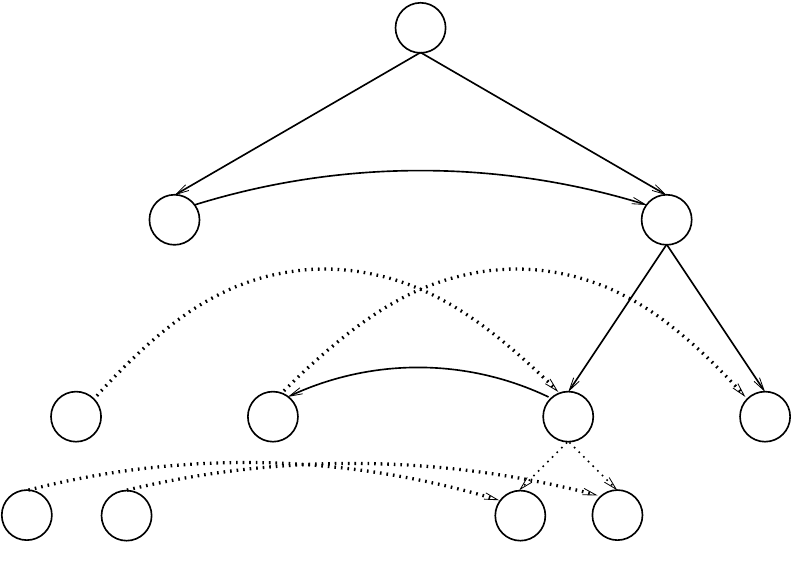}
  \caption{After two applications of the sum transformation} \label{twoapps}
\end{figure}

We see that only one application of the sum transformation can be
applied at level 1. At the next level we continue the process
begining with the new peak $x_{1}$. The result of applying the
sum transformation on $x_{1}$ is the removal of
$x_{1} \ueq x_{11}+x_{12}$ from the set of equations and the addition
of two new edges, $x_{11} \ueq T \times y_{11}$ and 
$x_{12} \ueq T \times y_{12}$, to the set of equations. 
The two new $y$ nodes are also
created adding $y_{1} \ueq y_{11}+y_{12}$ to the set of equations. The
result is that $x_{1}$ is no longer a peak but now $y_{1}$ is
(see Lemma~\ref{Lemma2}). The resulting graph can be seen in 
Figure~\ref{twoapps}.
Also, now that $y_{1}$ is the peak the direction of the
multiplication path has switched to the direction of $y_{1}$ to
$x_{2}$ (see Lemma~\ref{Lemma1}).

\begin{figure}[H]
  \centering
   \input{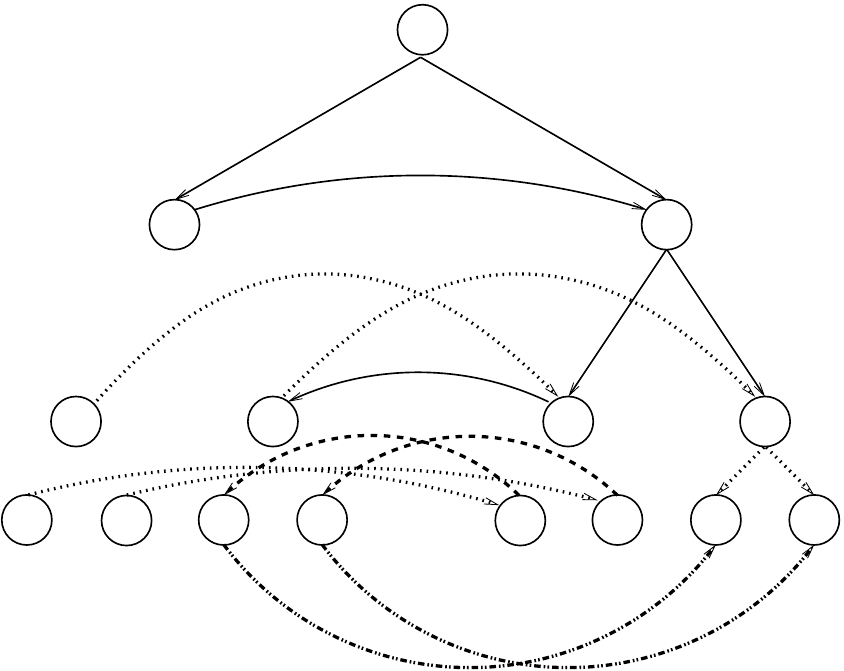}
  \caption{After 4 applications of the sum transformation} \label{fig6}
\end{figure}

We can now continue the process, applying the sum transformation
to the peak at $y_{1}$. This will remove $y_{1} \ueq y_{11}+y_{12}$
fom the set of equations and add $x_{2} \ueq x_{21}+x_{22}$ to the set
of equations, creating a peak at node $x_{2}$ and removing the
peak at node $y_{1}$. Lastly, a third sum transformation is
applied to node $x_{2}$, removing $x_{2} \ueq x_{21}+x_{22}$ from the
set of equations and adding $y_{2} \ueq y_{21}+y_{22}$ to the set of
equations. Note that because there is no multiplication path from
$y_{2}$ to some $x$ node $y_{2}$ is not a peak and no more sum
transformations can be applied at the current level. Because
there are also no additional nodes at the next level we stop with
a total of 4 applications of the sum transformation. The final
graph is shown in Figure~\ref{fig6}.

\section{Conclusions}

We have shown that the Tiden-Arnborg algorithm does not run in
polynomial time as claimed in~\cite{TidenArnborg87}. 
It is also not hard to see 
that the 
algorithm produces exponentially large
\textit{mgus} for the set of systems
$\sigma(n)$.  However, it may
still be that the {\em unifiability\/} problem,
i.e., whether a unifier exists modulo this theory, is in~\textsf{P}.
We are currently working on this and related problems.

\bibliographystyle{eptcs} 
\bibliography{report}

\nocite{BaaderSnyder,JouannaudKirchner91}

\end{document}